\documentclass[review,12pt]{elsarticle}
\usepackage{graphicx}
\usepackage{amssymb}

\journal{Physics Letters B}

\begin{document}

\begin{frontmatter}

\title{Experimental evidences of difference in $pp$ and $p\bar{p}$ interactions at high energies}

\author{V.A. Abramovsky\corref{cor1}}
\author{N.V. Radchenko}
\address{Novgorod State University, 173003 Novgorod the Great, Russia}
\cortext[cor1]{Victor.Abramovsky@novsu.ru}

\begin{abstract}
Hadrons production is different in $p\bar{p}$ and $pp$
interactions at high energies. There is process of hadrons
production from three quark strings in $p\bar{p}$ which is absent
in $pp$. This process grows as $(\ln\sqrt{s})^2$ and becomes
significant when energy of collision increases. Inclusive cross
sections of $p\bar{p}$ interaction exceed inclusive cross sections
of $pp$. Theoretical estimation of the ratio of $p\bar{p}$ to $pp$
at energy $\sqrt{s}=900$~GeV gives $R=1.12\pm0.03$. The UA1  data
on $p\bar{p}$ transverse momentum distribution are about 1.2 --
1.3 times higher than the CMS, ATLAS and ALICE  data on $pp$ at
energy $\sqrt{s}=900$~GeV.
\end{abstract}

\end{frontmatter}
\newpage
\section{Introduction}
The Collaborations CMS~\cite{bib1}, ATLAS~\cite{bib2} and
ALICE~\cite{bib3} have published inclusive charged particle
transverse momentum distributions in $pp$ interaction at
center-of-mass energy $\sqrt{s}=900$~GeV\footnote{We define
invariant inclusive cross section in standard form after pioneer
papers~\cite{bib4, bib5}
$E\frac{\mathrm{d}^3\sigma}{\mathrm{d}p^3}=\frac{1}{2\pi
p_T}\frac{\mathrm{d}^2\sigma}{\mathrm{d}y\,\mathrm{d}p_T}=\frac{1}{2\pi
p_T}\frac{E}{p}\frac{\mathrm{d}^2\sigma}{\mathrm{d}\eta\,\mathrm{d}p_T}$.
Here $E$, $p$ -- energy and momentum of observed particle, $p_T$
-- its transverse momentum, $y$ -- rapidity and $\eta$ --
pseudorapidity. Multiplicity density with respect to transverse
momentum in unit of rapidity
$\frac{\mathrm{d}^2n_{ch}}{\mathrm{d}y\,\mathrm{d}p_T}=\frac{1}{\sigma}\frac{\mathrm{d}^2\sigma}{\mathrm{d}y\,\mathrm{d}p_T}$
or pseudorapidity
$\frac{\mathrm{d}^2n_{ch}}{\mathrm{d}\eta\,\mathrm{d}p_T}=\frac{1}{\sigma}\frac{\mathrm{d}^2\sigma}{\mathrm{d}\eta\,\mathrm{d}p_T}$.
Value of $\sigma$ can be picked either as $\sigma=\sigma_{inel}$
(inelastic) or as $\sigma=\sigma_{NSD}$ (non single diffractive)
cross sections depending on various experimental methodologies.
Notations of ATLAS and ALICE therefore can be written as
$\frac{1}{N_{ev}}\frac{\mathrm{d}^2N_{ch}}{\mathrm{d}\eta\,\mathrm{d}p_T}=\frac{\mathrm{d}^2n_{ch}}{\mathrm{d}\eta\,\mathrm{d}p_T}$,
notation of CMS can be written as
$E\frac{\mathrm{d}^3N_{ch}}{\mathrm{d}p^3}=\frac{1}{\sigma}E\frac{\mathrm{d}^3\sigma}{\mathrm{d}p^3}$.}.
The ATLAS and ALICE compared their measurements to inclusive cross
sections of $p\bar{p}$ interaction obtained by the UA1
Collaboration~\cite{bib6} at the same energy $\sqrt{s}=900$~GeV.
The UA1 data overlaid with the ATLAS, ALICE and CMS data are shown
in Fig.1. As an immediate consequence of Fig.1,  the ratio of
$p\bar{p}$ to $pp$ inclusive cross sections at the same energy
$\sqrt{s}=900$~GeV
\begin{equation}\label{1}
R=\left[\frac{1}{2\pi
p_T}\frac{\mathrm{d}^2n_{ch}^{p\bar{p}}}{\mathrm{d}\eta\,\mathrm{d}p_T}\right]\Bigg/\left[\frac{1}{2\pi
p_T}\frac{\mathrm{d}^2n_{ch}^{pp}}{\mathrm{d}\eta\,\mathrm{d}p_T}\right]
\end{equation}
is greater than unity, $R\simeq1.2$ for the ATLAS and ALICE data
and $R\simeq1.3$ for the CMS data.
\begin{figure}[!h]
\centerline{
\includegraphics[scale=0.7]{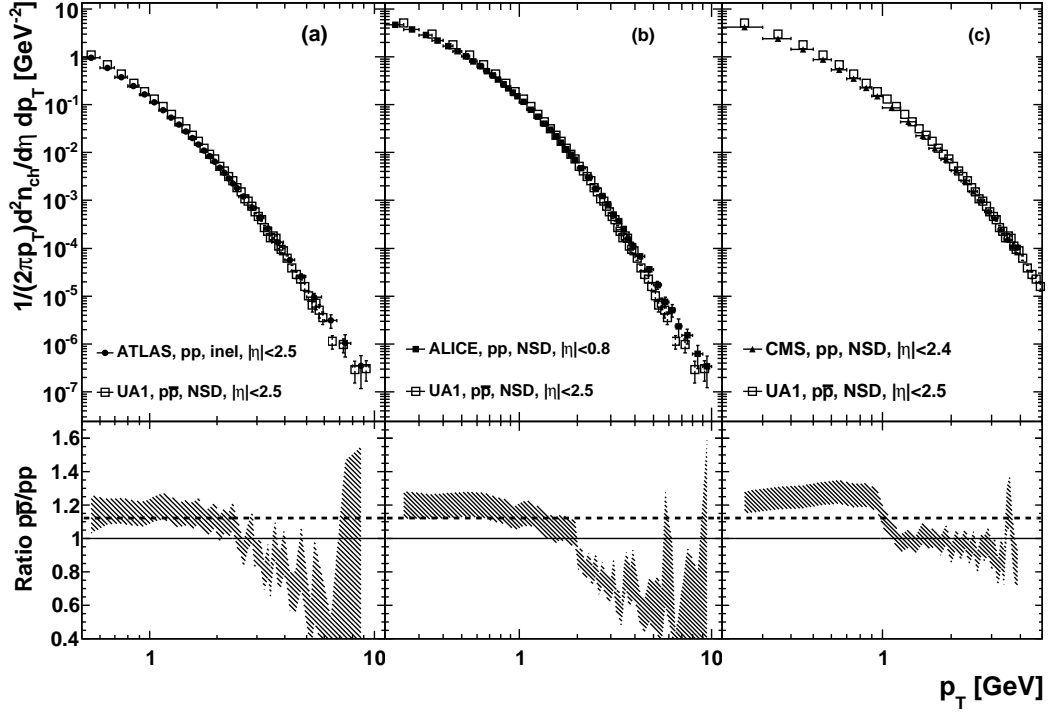}}\caption{The ratios of invariant inclusive cross
sections of the UA1~\cite{bib6} to ATLAS~\cite{bib2} (a),
ALICE~\cite{bib3} (b) and CMC~\cite{bib1} (c) at
$\sqrt{s}=900$~GeV. The shaded areas indicate the errors of the
ratios. The dashed line shows the value of ratio $R=1.12$, our
prediction from the LCNM. The solid line at unity is shown for
visibility.}
\end{figure}

The ATLAS and ALICE state that the difference is bound to
systematic uncertainties of the UA1 experiment. The CMS did not
compare their data on $p_T$ distribution with the UA1 data and
made no comments.

The Pomeranchuk theorem states that total, elastic and
differential elastic cross sections of $pp$ and $p\bar{p}$
interactions are equal at asymptotic high energies. It is commonly
believed that characteristics of multiple production such as, for
example, invariant inclusive cross section
$E\mathrm{d}^3\sigma/\mathrm{d}p^3$ are also equal for $pp$ and
$p\bar{p}$ interactions at high energies. So it is expected that
at high energies there must be equality $R=1$ and experimentalists
naturally try to explain the ratio $R>1$ by the UA1 uncertainties.

The purpose of the present work is to argue that inclusive cross
sections of $p\bar{p}$ interactions are higher than $pp$ at the
same energy. Thus experimental data of the CMS, ATLAS and ALICE do
correspond to reality.

\section{Inclusive cross sections of the CDF and CMS Collaborations}
Our first argument is based on analysis of the CDF data on
$p\bar{p}$ interactions at $\sqrt{s}=1.96$~TeV~\cite{bib7}
together with the CMS data on $pp$ interactions at
$\sqrt{s}=2.36$~TeV~\cite{bib1}. We calculated the ratio of
$E\mathrm{d}^3\sigma^{p\bar{p}}/\mathrm{d}p^3$ to
$E\mathrm{d}^3\sigma^{pp}/\mathrm{d}p^3$ which is shown in Fig.2.
The result is amazing -- the ratio of inclusive cross sections
equals unity with good accuracy. If we accept that
$E\mathrm{d}^3\sigma^{p\bar{p}}/\mathrm{d}p^3=E\mathrm{d}^3\sigma^{pp}/\mathrm{d}p^3$
at the same energy, these cross sections must be different as the
energy increases by 400~GeV. That is,
$E\mathrm{d}^3\sigma^{pp}/\mathrm{d}p^3$ at $\sqrt{s}=2.36$~TeV
must be higher than $E\mathrm{d}^3\sigma^{p\bar{p}}/\mathrm{d}p^3$
at $\sqrt{s}=1.96$~TeV. Therefore the ratio given in the lower
panel of Fig.2 must be systematically lower than unity.

\begin{figure}[!h]
\centerline{
\includegraphics[scale=0.7]{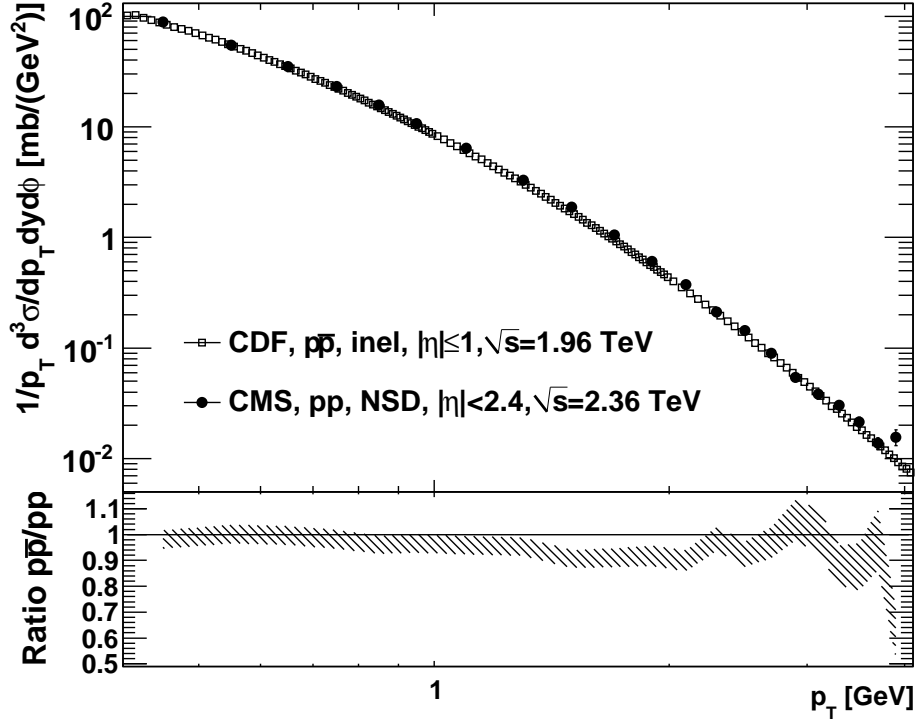}}\caption{The ratio of invariant inclusive cross
sections of CDF~\cite{bib7} at $\sqrt{s}=1.96$~TeV to
CMS~\cite{bib1} at $\sqrt{s}=2.36$. The shaded area indicates the
errors of the ratio. The solid line at unity is shown for
visibility.}
\end{figure}

Let us discuss some details of our analysis. Since there are no
measurements of $p\bar{p}$ cross sections (total, inelastic or
NSD) at $\sqrt{s}=1.96$~TeV, we obtained the inclusive cross
section $E\mathrm{d}^3\sigma^{pp}/\mathrm{d}p^3$ from the CMS data
$\frac{1}{2\pi
p_T}\frac{\mathrm{d}^2N_{ch}}{\mathrm{d}\eta\,\mathrm{d}p_T}$~\cite{bib1}
multiplied by $\sigma_{NSD}$. We used value
$\sigma_{NSD}=49.86$~mb which was estimated by the ALICE for
$\sqrt{s}=2.36$~TeV~\cite{bib8}. (Lower value
$\sigma_{NSD}=48.77$~mb which gives lower value of $pp$ inclusive
cross section was obtained in~\cite{bib9}.) Therefore the
experimental ratio given in Fig.~2 presumably confirms our
assumption that
$E\mathrm{d}^3\sigma^{p\bar{p}}/\mathrm{d}p^3>E\mathrm{d}^3\sigma^{pp}/\mathrm{d}p^3$
at the same energy.

\section{Subprocesses of multiple production in $pp$ and $p\bar{p}$}

In this section we will argue that it is possible to explain the
difference in inclusive spectra of $pp$ and $p\bar{p}$
interactions. Previously we have demonstrated the possibility of
difference in multiplicity distributions in $pp$ and $p\bar{p}$
scatterings~\cite{bib10}. It is almost impossible to prove this
difference experimentally. But we can use this approach to analyze
possible inequality in
$E\mathrm{d}^3\sigma^{p\bar{p}}/\mathrm{d}p^3$ and
$E\mathrm{d}^3\sigma^{pp}/\mathrm{d}p^3$. We are based on the Low
Constituents Number Model (LCNM)~\cite{bib11, bib12}. As it should
be in any collision theory, the model contains three stages:
preparation of initial state, interaction, and separation of
reaction products. Schematic illustration of hadrons interaction
and multiple production in this model is given by phenomenological
diagrams in Fig.3.
\begin{figure}[!h]
\centerline{
\includegraphics[scale=0.7]{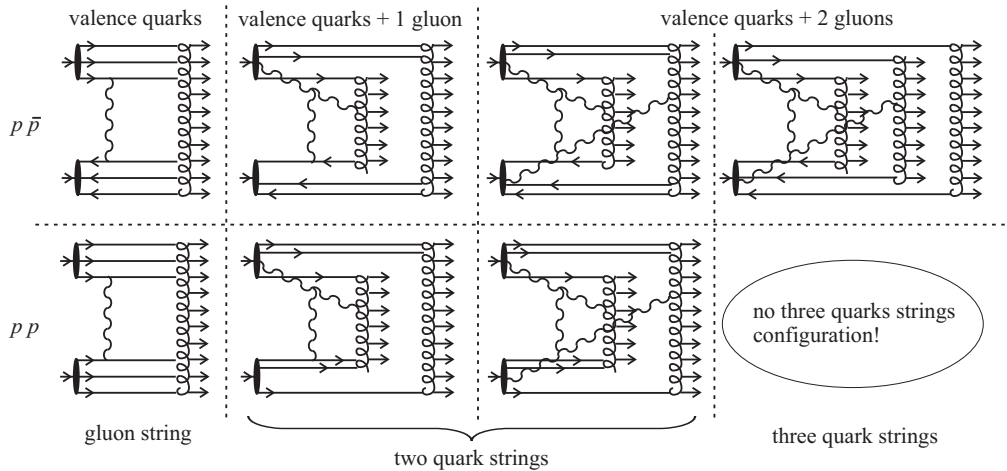}}
\caption{Three types of inelastic subprocesses in $p\bar{p}$ and
$pp$ scattering in the LCNM. Solid lines correspond to valence
quarks and antiquarks, wavy lines are gluons. Color field string
is shown as spiral. The initial state can be composed of either
only valence quarks or valence quarks plus 1 or 2 gluons.  The
final state is different for $pp$ and $p\bar{p}$ -- there is no
three quark strings configuration in $pp$ interaction.}
\end{figure}

A key feature of the model is assumption that there are only few
scatters (constituents) in initial state of each colliding hadrons
-- these are valence quarks (antiquarks) and only one gluon. This
gluon appears with low probability which grows slowly as energy
increases\footnote{This assumption allows to explain small value
of the Pomeranchuk trajectory slope~\cite{bib11}. It also allows
to describe value and growth with energy of $pp$, $p\bar{p}$,
$\pi^\pm p$, $K^\pm p$ total cross sections~\cite{bib12}.}.
Interaction is carried out by gluon exchange which corresponds to
Low--Nussinov two-gluon pomeron~\cite{bib13}. Due to gluon
exchange colorless hadrons gain color charge. Separation of
reaction products (colored hadrons) occurs after interaction. When
the charges are separated by distance larger than the confinement
radius, color electric field strings are produced. These strings
break up to primary hadrons. Formation of color field string and
its breakup into primary hadrons corresponds to the interaction in
the final state.

It should be emphasized that because color objects are not
emitted, the process of converting color charges to hadrons occurs
with probability which equals to unity and does not affect
interaction probability. Therefore values of total cross sections
for both $pp$ and $p\bar{p}$ are defined only by one gluon
exchange in the second stage and so they are the same. Thus the
Pomeranchuk theorem is fulfilled in the proposed LCNM approach.

We distinguish the following inelastic processes of hadrons
production (or they might be better defined as inelastic
subprocesses in multiple hadron production) in $pp$ and $p\bar{p}$
interactions.

\emph{Hadrons production from decay of gluon string.} Gluon string
is produced when objects carrying octet quantum numbers fly apart
after interaction. In this case it is impossible to separate gluon
from valence quarks. Wavelength of the gluon is such, that it
overlaps with the valence quarks. This subprocess gives constant
contribution to total cross sections. This subprocess is the same
for both $pp$ and $p\bar{p}$ interactions (Fig.3, first column).

\emph{Hadrons production from decay of two quark strings.} Quark
strings are produced between quark and antiquark and between
diquark and antidiquark in $p\bar{p}$ interaction and between
quark and diquark in $pp$ interaction. Since gluon spectrum is
$\mathrm{d}\omega/\omega$ ($\omega$ -- gluon energy), contribution
from the component with one gluon in the initial state grows as
$\ln\sqrt{s}$. This gluon is absorbed after the interaction by one
of the quark strings, and it changes the string color charge --
``recolor'' the string. This contribution is the same for both
$pp$ and $p\bar{p}$ interactions (Fig.3, second column).

The contribution from two gluons in the initial state grows as
$(\ln\sqrt{s})^2$. Both gluons are absorbed by quark strings
``recoloring'' them. In case of $pp$ interaction the two gluons
initial state can only give configuration with two quark strings
in the final state (Fig.3, third column).

\emph{Hadrons production from decay of three quark strings.} In
case of $p\bar{p}$ interaction the two gluons initial state
besides the configuration with two quark strings can lead to
configuration with three quark strings (Fig.3, fourth column). The
quark strings are produced between each  quark and each antiquark.
Since the contribution of this subprocesses grows as
$(\ln\sqrt{s})^2$, it is quite essential at high
energies\footnote{Our approach is different from approach with
exchange of decameron~\cite{bib14}, which gives difference in
multiplicity distributions and inclusive cross sections in  $pp$
and $p\bar{p}$ interactions. Contribution from decameron is low
and constant with energy~\cite{bib14}. The AFS~\cite{bib15} and
UA5~\cite{bib16} Collaborations have not found this contribution
at energy $\sqrt{s}=53$~GeV. Moreover it will not be seen at
higher energies.}.

We suppose that difference in inclusive cross sections in $pp$ and
$p\bar{p}$ interactions is governed by presence of three quark
strings in $p\bar{p}$. This subprocess gives contribution in
multiplicity distribution $P_n$ in the region of high $n$ ($n$ -
number of charged particles). In this region the value of $P_n$ is
about one order of magnitude smaller than in the maximum region
and so it is hard to experimentally study it because of large
uncertainties. Therefore  the difference in $P_n$ distributions in
$pp$ and $p\bar{p}$ interactions will be difficult to detect. One
has to use another observable value, which is able to increase the
difference between $P_n$ distributions  in $pp$ and $p\bar{p}$
scatterings. We propose to use $nP_n$ as the required observable
value.

Let us define inclusive cross section of one charged particle
production in an event with $n$ charged particles -- a topological
inclusive cross section (``semi-inclusive'' cross section of Koba,
Nielsen and Olesen~\cite{bib17})
\begin{equation}\label{2}
E\frac{\mathrm{d}^3\sigma_n}{\mathrm{d}p^3},\;\;\;\;\int\mathrm{d}p^3\frac{\mathrm{d}^3\sigma_n}{\mathrm{d}p^3}=n\sigma_n,
\end{equation}
where $\sigma_n$ -- topological cross section of $n$ charged
particles production. We consider here only non single diffractive
events, so $\sum_n\sigma_n=\sigma_{NSD}$. We stress that (\ref{2})
is normalized to $n\sigma_n$, where $n$ -- number of particles in
an event.

In what follows we are based on the UA5 Collaboration
data~\cite{bib18} on the inclusive cross sections in 9
multiplicity bins: $2\leqslant n\leqslant 10$, $12\leqslant n
\leqslant20$, \ldots, $n\geqslant 82$ at energy
$\sqrt{s}=900$~GeV. We define inclusive cross sections in bin
($i$)
\begin{equation}\label{3}
\frac{\mathrm{d}^3\sigma^{(i)}}{\mathrm{d}p^3}=\sum_{n\mbox{\scriptsize{
in bin }}(i)}\frac{\mathrm{d}^3\sigma_n}{\mathrm{d}p^3}
\end{equation} which are
normalized as follows
\begin{equation}\label{4}
\int\mathrm{d}p^3\frac{\mathrm{d}^3\sigma^{(i)}}{\mathrm{d}p^3}=\sigma_{NSD}\sum_{n\mbox{\scriptsize{
in bin }}(i)}nP_n=\sigma_{NSD}\,\bar{n}^{(i)}.
\end{equation}
Here $P_n=\sigma_n/\sigma_{NSD}$ -- probability of $n$ charged
particles production in a NSD event. Since we belive that
inclusive cross sections of $pp$ and $p\bar{p}$ are different we
write down relation~(\ref{4}) separately for $pp$ and $p\bar{p}$.
It was shown in~\cite{bib19} that single diffractive cross
sections $\sigma_{SD}$ are the same for $pp$ and $p\bar{p}$
interactions. Therefore
$\sigma_{NSD}=\sigma_{tot}-\sigma_{el}-\sigma_{SD}$ are also the
same.

From ratio of $pp$ to $p\bar{p}$ in (\ref{4}) we obtain the
following relation
\begin{equation}\label{5}
\int\mathrm{d}p^3\frac{\mathrm{d}^3\sigma^{(i)}_{pp}}{\mathrm{d}p^3}=
\frac{\bar{n}^{(i)}_{pp}}{\bar{n}^{(i)}_{p\bar{p}}}\int\mathrm{d}p^3\frac{\mathrm{d}^3\sigma^{(i)}_{p\bar{p}}}{\mathrm{d}p^3}.
\end{equation}
Value of $\bar{n}^{(i)}_{pp}/\bar{n}^{(i)}_{p\bar{p}}$ does not
depend on momentum of observed particle $p$. Besides, bin limits
can be chosen arbitrary. Therefore one of solutions of (\ref{5})
(perhaps, the only solution) has the form
\begin{equation}\label{6}
\frac{\mathrm{d}^3\sigma^{(i)}_{pp}}{\mathrm{d}^3p}=
\frac{\bar{n}^{(i)}_{pp}}{\bar{n}^{(i)}_{p\bar{p}}}\,\frac{\mathrm{d}^3\sigma^{(i)}_{p\bar{p}}}{\mathrm{d}^3p}.
\end{equation}
(If $pp$ and $p\bar{p}$ interactions are the same, we obtain a
trivial result.) The relation for the inclusive pseudorapidity
cross sections in bin $(i)$ is
\begin{equation}\label{7}
\frac{\mathrm{d}\sigma^{(i)}_{pp}}{\mathrm{d}\eta}=
\frac{\bar{n}^{(i)}_{pp}}{\bar{n}^{(i)}_{p\bar{\bar{p}}}}\,\frac{\mathrm{d}\sigma^{(i)}_{p\bar{p}}}{\mathrm{d}\eta}.
\end{equation}

We calculated the values of $\bar{n}^{(i)}_{p\bar{p}}$ from the
UA5 data in each bin. The values of $\bar{n}^{(i)}_{pp}$ are
calculated from multiplicity distribution $P_n^{pp}$, obtained in
frame of LCNM~\cite{bib10}. Then we obtained inclusive
pseudorapidity cross section for $pp$ interaction at
$\sqrt{s}=900$~GeV.
\begin{equation}\label{8}
\frac{\mathrm{d}\sigma_{pp}}{\mathrm{d}\eta}=\sum_i\frac{\mathrm{d}\sigma^{(i)}_{pp}}{\mathrm{d}\eta}=\sum_{i=1}^9
\frac{\bar{n}^{(i)}_{pp}}{\bar{n}^{(i)}_{p\bar{p}}}\,\frac{\mathrm{d}\sigma^{(i)}_{p\bar{p}}}{\mathrm{d}\eta}
\end{equation}
The values of the inclusive cross sections
$\mathrm{d}\sigma_{p\bar{p}}/\mathrm{d}\eta$ and
$\mathrm{d}\sigma_{pp}/\mathrm{d}\eta$ are shown in Fig.4 for
whole multiplicity range (a) and for multiplicity from 62 to 70
(b). In this multiplicity  bin the difference in the inclusive
cross sections is very large because of high values of $n$.

\begin{figure}[!h]
\centerline{
\includegraphics[scale=0.7]{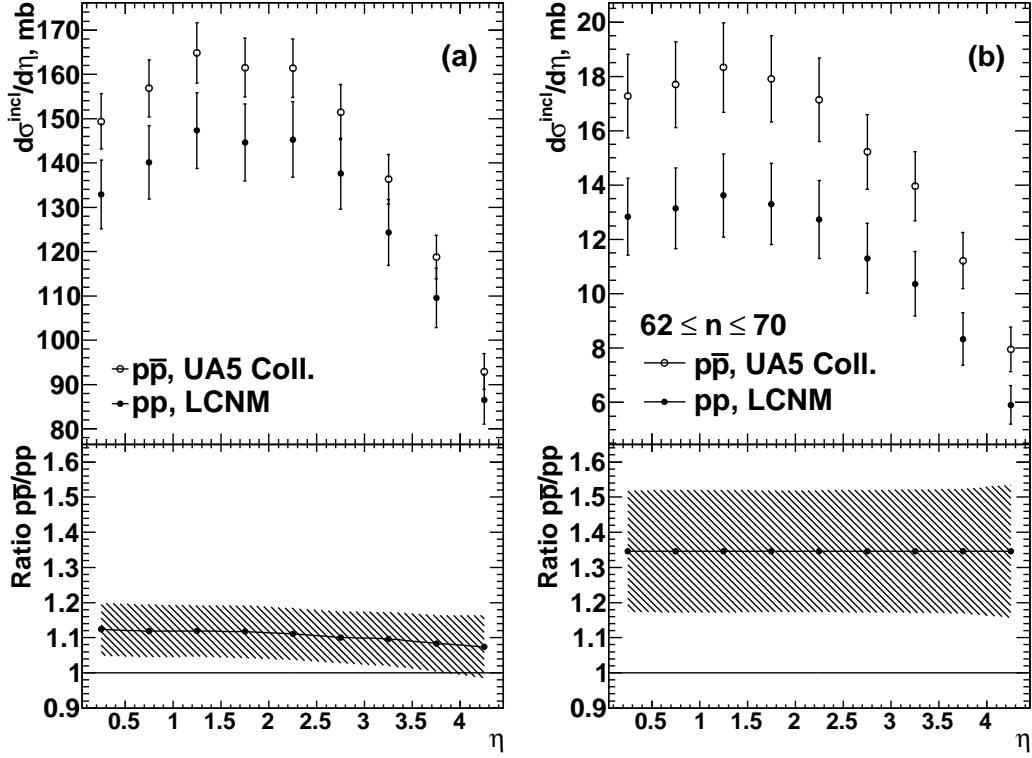}}
\caption{Inclusive cross sections at $\sqrt{s}=900$~GeV. The
points for $p\bar{p}$ were obtained from the UA5
data~\cite{bib18}, the points for $pp$ were obtained from the
calculations in the LCNM. (a) Inclusive cross section for all
charged particles. (b) Inclusive cross section for charged
multiplicity bin $62\leqslant n \leqslant 70$.}
\end{figure}

Factorization of inclusive cross sections results from the
Abramovsky--Gribov--Kacheli (AGK cancellations)
theorem~\cite{bib20}. Phenomenological factorization relations
were proposed by Hagedorn~\cite{bib21} and Tsallis~\cite{bib22}.
Therefore we can write down separate formulas for $pp$ and
$p\bar{p}$
\begin{equation}\label{9}
\frac{1}{2\pi
p_{\mathrm{T}}}\frac{\mathrm{d}^2\sigma_{pp}}{\mathrm{d}\eta\,\mathrm{d}p_{\mathrm{T}}}=
f_{pp}(p_{\mathrm{T}})\frac{\mathrm{d}\sigma_{pp}}{\mathrm{d}\eta},
\quad\frac{1}{2\pi
p_{\mathrm{T}}}\frac{\mathrm{d}^2\sigma_{p\bar{p}}}{\mathrm{d}\eta\,\mathrm{d}p_{\mathrm{T}}}=
f_{p\bar{p}}(p_{\mathrm{T}})\frac{\mathrm{d}\sigma_{p\bar{p}}}{\mathrm{d}\eta}.
\end{equation}

It is easy to obtain from the relations~(\ref{5}) -- (\ref{9})
that $f_{pp}(p_{\mathrm{T}})\equiv f_{p\bar{p}}(p_{\mathrm{T}})$.
One can obtain from~(\ref{9}) the following ratio
\begin{equation}\label{10}
\left(\frac{1}{2\pi
p_{\mathrm{T}}}\frac{\mathrm{d}^2\sigma_{p\bar{p}}}{\mathrm{d}\eta\,\mathrm{d}p_{\mathrm{T}}}\right)\left/
\left(\frac{1}{2\pi
p_{\mathrm{T}}}\frac{\mathrm{d}^2\sigma_{pp}}{\mathrm{d}\eta\,\mathrm{d}p_{\mathrm{T}}}\right)\right.=
\frac{\mathrm{d}\sigma_{p\bar{p}}}{\mathrm{d}\eta}\left/\frac{\mathrm{d}\sigma_{pp}}{\mathrm{d}\eta}\right.=R.
\end{equation}
It should be noted that the relations~(\ref{9}), (\ref{10}) must
be fulfilled in region of soft physics where the AGK theorem is
valid. Therefore the relation~(\ref{10}) must be fulfilled  for
transverse momenta up to $p_T=1.5\div2$~GeV. As it can be seen in
Fig.4, the ratio of the inclusive cross sections is approximately
equal to $R\simeq1.12$. We can write down more strictly
$R=1.12\pm0.03$~\cite{bib23}.  On basis of relation~(\ref{10}) we
can state that ratio of inclusive cross sections of $p\bar{p}$ to
$pp$ is equal to $1.12\pm0.03$. This value is depicted in Fig.1 by
dashed line.

\section{Discussion}
We have shown that excess of $p\bar{p}$ inclusive cross section
over $pp$ inclusive cross section at the same energy is connected
with hadrons production in three quarks string configuration in
$p\bar{p}$, which is absent in $pp$ interaction. We have predicted
this difference in our paper~\cite{bib23} before the data of the
CMS~\cite{bib1}, ATLAS~\cite{bib2} and ALICE~\cite{bib3} were
published. It should be noted that difference in $pp$ and
$p\bar{p}$ increases with rising collision energy in our
approach\footnote{Three-sheet annihilation of
Rossi--Veneziano~\cite{bib24} decreases as $s^{-1/2}$.}. We
emphasize that we have a physical picture which was, probably,
confirmed by results of the experiments UA1, CDF, CMS, ATLAS and
ALICE.

We do not agree with the following statements of the ALICE and
ATLAS. ``The excess of the UA1 data of about 20\% at low $p_T$ is
possibly due to the UA1 trigger condition, which suppresses events
with very low multiplicity''~\cite{bib3}. ``A shift in this
direction is expected from the double-arm scintillator trigger
requirement used to collect the UA1 data, which rejected events
with low charged-particle multiplicities''~\cite{bib2}.

Inclusive cross sections in our notations and in notations of the
ATLAS and ALICE can be written as equality
\begin{equation}\label{11}
\frac{1}{2\pi
p_T}\frac{\mathrm{d}^2n_{ch}}{\mathrm{d}\eta\mathrm{d}p_T}=\frac{1}{N_{ev}}\frac{1}{2\pi
p_T}\frac{\mathrm{d}^2N_{ch}}{\mathrm{d}\eta\mathrm{d}p_T}
\end{equation}
where $N_{ev}$ is the number of events inside the selected
kinematic range, $N_{ch}$ is the total number of charged particles
in the data sample. It follows from simple logic that increase of
inclusive cross section in the left part of~(\ref{11}) is possible
if $N_{ev}$ decreases, and that is the basis of the ATLAS and
ALICE statements. However, it is very hard to accept that the UA1
Collaboration have lost about 17\% of their data in case of
comparing with the ATLAS and ALICE and 21\% in case of comparing
with the CMS.

In this case a serious discrepancy arises, which should be noticed
by experimentalists. The Hagedorn--Tsallis factorization formula
is generally accepted by experimentalists for describing of
transverse momentum dependence of inclusive cross sections. Let us
just use it to data given in Fig.1 without pointing out any
theoretical considerations. Since all data were taken at the same
energy $\sqrt{s}=900$~GeV then all factors determining $p_T$
dependence cancel each other. Therefore it follows from data given
in Fig.1 that
$\mathrm{d}n_{ch}^{p\bar{p}}/\mathrm{d}\eta>\mathrm{d}n_{ch}^{pp}/\mathrm{d}\eta$.
From the other side, direct measurement of the CMS~\cite{bib1} and
ALICE~\cite{bib25} gave
$\mathrm{d}n_{ch}^{p\bar{p}}/\mathrm{d}\eta\approx\mathrm{d}n_{ch}^{pp}/\mathrm{d}\eta$.

We want to stress that discovery of an additional inelastic
process in $p\bar{p}$ interaction is very important by itself. If
existence of this process is experimentally proved, it will
greatly change theoretical concepts of high energy physics. It is
also important for Monte Carlo event generators which use data on
$pp$ and $p\bar{p}$ simultaneously in tuning of parameters, what
may produce incorrect results.

\section{Acknowledgements}
We would like to thank O.V.~Kancheli for useful discussions  and
I.I.~Tsukerman for information support. We thank A.V.~Dmitriev for
discussions. V.A.A. acknowledges financial support by grant of
RFBR 11-02-01395-a. N.V.R.  acknowledges financial support by
grant of Ministry of education and science of the Russian
Federation, federal target program ``Scientific and
scientific-pedagogical personnel of innovative Russia'', grant
P1200.

\bibliographystyle{model1-num-names}

\begin{thebibliography}{00}
\bibitem{bib1}
  V.~Khachatryan et al. CMS Collaboration,
  JHEP  1002 (2010) 041,
  arXiv:1002.0621 [hep-ex].

\bibitem{bib2}
  G.~Aad et al.  ATLAS Collaboration,
  Phys. Lett. B  688 (2010) 21,
  arXiv:1003.3124 [hep-ex].

\bibitem{bib3}
  K.~Aamodt et al.  ALICE Collaboration,
  Phys. Lett. B 693 (2010) 53,
  arXiv:1007.0719 [hep-ex].

\bibitem{bib4}
  A.~H.~Mueller,
  Phys. Rev. D 2 (1970) 2963.

\bibitem{bib5}
  O.~V.~Kancheli,
  Pisma Zh. Eksp. Teor. Fiz. 11 (1970) 397.

\bibitem{bib6}
  C.~Albajar et al. UA1 Collaboration,
  Nucl. Phys. B 335 (1990) 261.

\bibitem{bib7}
  T.~Aaltonen et al. CDF Collaboration,
  Phys. Rev.  D  79 (2009) 112005,
  arXiv:0904.1098 [hep-ex].

\bibitem{bib8}
  M.~G.~Poghosyan for ALICE Collaboration,
  J. Phys. G 38 (2011) 124044,
  arXiv:1109.4510 [hep-ex].

\bibitem{bib9}
  J.~Bleibel, L.~V.~Bravina, A.~B.~Kaidalov, E.~E.~Zabrodin,
  arXiv:1011.2703 [hep-ph].

\bibitem{bib10}
  V.~A.~Abramovsky, N.~V.~Radchenko,
  Phys. Part. Nucl. Lett. 6 (2009) 433.

\bibitem{bib11}
  V.~A.~Abramovsky, O.~V.~Kancheli,
  Pisma Zh. Eksp. Teor. Fiz. 31 (1980) 566;

  V.~A.~Abramovsky, O.~V.~Kancheli,
  Pisma Zh. Eksp. Teor. Fiz. 32 (1980) 498.

\bibitem{bib12}
  V.~A.~Abramovsky, N.~V.~Radchenko,
  Phys. Part. Nucl. Lett.  6 (2009) 368.

\bibitem{bib13}
  F.~E.~Low,
  Phys. Rev.  D 12 (1975) 163;

  S.~Nussinov,
  Phys. Rev. Lett. 34 (1975) 1286.

\bibitem{bib14}
  B.~Z.~Kopeliovich, B.~G.~Zakharov,
  Phys. Lett.  B 211 (1988) 221.

\bibitem{bib15}
  T.~Akesson et al. Axial Field Spectrometer Collaboration,
  Phys. Lett.   B 108 (1982) 58.

\bibitem{bib16}
  K.~Alpgard  et al. UA5 Collaboration,
  Phys. Lett.  B 112 (1982) 183.

\bibitem{bib17}
  Z.~Koba, H.~B.~Nielsen, P.~Olesen,
  Phys. Lett.   B 38 (1972) 25.

\bibitem{bib18}
  G.~J.~Alner et al. UA5 Collaboration,
  Z. Phys. C 33 (1986) 1.

\bibitem{bib19}
  V.~A.~Abramovsky,
  arXiv:0911.4850 [hep-ph].

\bibitem{bib20}
  V.~A.~Abramovsky, V.~N.~Gribov, O.~V.~Kancheli,
  Yad. Fiz. 18 (1973) 595.

\bibitem{bib21}
  R.~Hagedorn,
  Riv. Nuovo Cim.  6N10 (1984) 1.

\bibitem{bib22}
  C.~Tsallis,
  J. Stat. Phys. 52 (1988) 479.

\bibitem{bib23}
  V.~A.~Abramovsky, N.~V.~Radchenko,
  arXiv:0912.1041 [hep-ph].

\bibitem{bib24}
  G.~C.~Rossi, G.~Veneziano,
  Nucl. Phys.   B 123 (1977) 507.

\bibitem{bib25}
  K.~Aamodt et al. ALICE Collaboration,
  Eur. Phys. J.  C 65 (2010) 111.
  arXiv:0911.5430 [hep-ex].

\end{thebibliography}

\end{document}